\documentclass[10pt,twocolumn,letterpaper]{article}

\usepackage[pagenumbers]{cvpr} 

\usepackage{booktabs}
\usepackage{tabularx}
\usepackage{multirow}
\usepackage{colortbl}
\usepackage{arydshln} 
\usepackage{epstopdf}

\makeatletter
\renewcommand{\cvprssubsubsect}[1]{%
  \@startsection{subsubsection}{3}{\z@}%
    {-6pt plus -2pt minus -2pt}%
    {-1em}%
    {\tenbf}*{#1}%
}
\makeatother

\usepackage{microtype}

\renewcommand{\paragraph}[1]{\vspace{0em}\noindent\textbf{#1.}}

\definecolor{cvprblue}{rgb}{0.21,0.49,0.74}
\usepackage[pagebackref,breaklinks,colorlinks,allcolors=cvprblue]{hyperref}

\def\confName{CVPR}
\def\confYear{2027}

\title{Your Secrets Are Showing: \\ Deep Image Hiding Models Are Surprisingly Easy to Attack}

\author{
  Hao Liu$^{1,2}$ \quad
  Jing Yu$^{3}$ \quad
  Weizhi Liu$^{1}$ \quad
  Lingling Yuan$^{1}$,\\
  Fengyong Li$^{4}$ \quad
  Keke Gai$^{5}$ \quad
  Zhaoxia Yin$^{1}$\setcounter{footnote}{1}%
\thanks{Corresponding author.} \quad
  Xinpeng Zhang$^{6}$\\[10pt]
  {\normalsize
    $^{1}$East China Normal University \quad
    $^{2}$Zhongguancun Academy \quad
    $^{3}$Minzu University of China
  }\\
  {\normalsize
    $^{4}$Shanghai University of Electric Power \quad
    $^{5}$Beijing Institute of Technology \quad
    $^{6}$Shanghai University
  }
  \vspace{-0.5cm}
}

\makeatletter
\apptocmd{\@maketitle}{%
  \par
  \noindent
  \begin{minipage}{\textwidth}
    \centering
    \includegraphics[width=\linewidth]{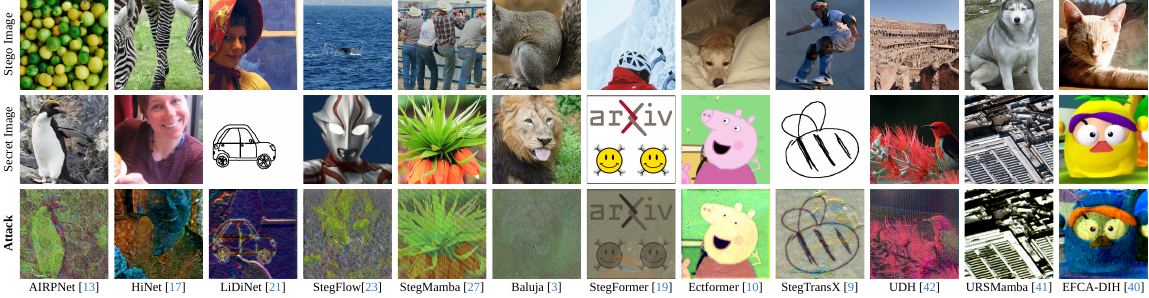}
    \captionsetup{hypcap=false}
    \captionof{figure}{%
    \textbf{Is deep image hiding or steganography truly secure?}
    We show that an attacker can use \emph{only one secret-recovery decoder} (here, we use HiNet~\cite{HiNet}, a well-known, open-source hiding model) to \emph{recover semantically meaningful secret images} (third row) from stego images (first row) generated by \emph{unseen hiding models with diverse architectures} (remaining columns, \emph{e.g.}, CNNs, Transformers, and Mamba), in a \emph{zero-shot} manner without target-specific training or fine-tuning. This capability extends to out-of-distribution secret images, as the results for \emph{sketches, logos, and cartoons} show.
    }
    \label{fig:intro1}
  \end{minipage}
  \par
  \vspace{1em}
}{}{%
  \PackageError{teaser}{Failed to patch \string\@maketitle}{}
}
\makeatother

\begin{document}
\maketitle
\begin{abstract}
In this work, we ask whether an attacker can go beyond detecting a hidden image and tell what it depicts, even without knowing which model hid it. We show that, for deep image hiding, this is surprisingly easy. To test this, we build an evaluation dataset of stego images generated by 11 hiding models. These models span two families (invertible neural networks and autoencoders), use different backbones (CNN, Transformer, and Mamba), and include both spatial- and frequency-domain hiding methods. We demonstrate that training HiNet~\cite{HiNet} in the usual way yields a single decoder that can extract secret content from unseen models across these settings. Both humans and machines can recognize what it reveals. A frequency analysis shows that most models favor mid- and high-frequency embedding. Guided by this finding, we strengthen cross-model extraction by enhancing cover textures and suppressing secret textures during training. The attack extends to out-of-distribution secrets and to multi-image hiding. It even exposes private content hidden by AIRPNet~\cite{AIRPNet}, a recent image restoration method that uses steganography for privacy protection. What surprised us most is that decoders from every hiding model we test exhibit this ``universal'' extraction ability to varying degrees. Our code and datasets will be open-sourced.
\end{abstract}    
\section{Introduction}
\label{sec:introduction}

Steganography aims to hide the existence of secret information by embedding it in an innocent carrier, enabling covert communication over public channels~\cite{steg-intro, zhu2018hidden}. In image steganography, the carrier is a \emph{cover image}, and the image containing the hidden information is called a \emph{stego image}. Traditional methods usually embed a relatively small payload, aiming to preserve the cover's visual appearance and avoid detection by steganalysis while allowing the intended receiver to recover the hidden data~\cite{LSB,STC-arch,ASDLGAN}. A central challenge is to increase this payload without compromising these goals~\cite{Digital-image-steganography}. \textbf{Image hiding} or \textbf{image-in-image steganography} takes this challenge further by hiding an entire \emph{secret image} within a cover image. It seeks to preserve the cover's appearance while allowing the secret image to be recovered with high visual quality.

An early milestone was Baluja's use of deep neural networks (DNNs) to implicitly learn representations for image hiding~\cite{baluja_nips, baluja2019hiding}. Later, HiNet~\cite{HiNet} used invertible neural networks (INNs) to model hiding and recovery as the forward and inverse processes of the same network~\cite{Glow}. These advances have also opened up new ways to \textbf{protect sensitive visual content}. The idea is to hide such \emph{sensitive content} inside an \emph{innocent cover image} and allow a downstream network to perform tasks on the hidden content by directly processing the stego image. For example, AIRPNet~\cite{AIRPNet} performs image restoration (\emph{e.g.}, denoising, deblurring, and super-resolution) on hidden secret images by processing their stego images. StegaVAR~\cite{stegavar} explores video action recognition in the steganographic domain.

Visual concealment, however, does not guarantee security. Previous studies have shown that image-in-image steganography is often vulnerable to steganalysis~\cite{iSCMIS}. An attacker may need as few as 50 cover-stego image pairs from a target model to train an effective detector~\cite{PUSNet}. Recent work further shows that such detectors can generalize, to some extent, to models unseen during training~\cite{ZS-Steganalysis}. This means that once an attacker detects a covert exchange, they can block the transmission or even keep copies of the stego images being sent. Nevertheless, the sender might still expect the hidden content to remain private. After all, \emph{a binary detector can tell the attacker whether a secret is present, but it cannot tell them what that secret depicts.} This distinction raises a more direct question: \textbf{Can an attacker go beyond detection and extract the hidden content itself}, even without any additional information about the target model (\emph{i.e.}, in a \emph{zero-shot} manner)? If so, the privacy protection offered by these applications would be directly compromised.

In this work, we show that hidden content \textbf{can indeed} be extracted with surprisingly little effort. An attacker trains a single image hiding model \emph{in the usual way} on a dataset of their choice, without access to the target models or their stego images during training. They then freeze their decoder and apply it directly to stego images from these \emph{unseen} models. As shown in~\cref{fig:intro1}, the cover still shows through in the extracted images, yet we can already tell what the secret images depict. The stego images may still look innocent, but under the same decoder, \emph{their secrets start to show}.

Why can one decoder recover secrets hidden by different models, and what helps it generalize? To find out, we build an evaluation dataset of 83,050 stego images generated by 11 hiding models and compare each image with its corresponding cover. These models span multiple families, backbone architectures, and embedding domains. A frequency analysis shows that most of these models favor mid- and high-frequency embedding. Guided by this finding, we introduce two \emph{simple yet effective} augmentation settings that an attacker can readily use during training to make the leaked content clearer. We also find that an attacker need not use a specialized hiding model. If they simply use a standard ResNet~\cite{resnet} to build the hiding and recovery networks, the trained decoder still reveals faint traces of the secret. Our extensive cross-model extraction experiments and a closer analysis of decoder channel responses show that a decoder trained to recover secrets hidden by its own encoder can also recognize embedding patterns shared across hiding models.

\begin{figure}[!t]
\begin{center}
\includegraphics[width=\columnwidth]{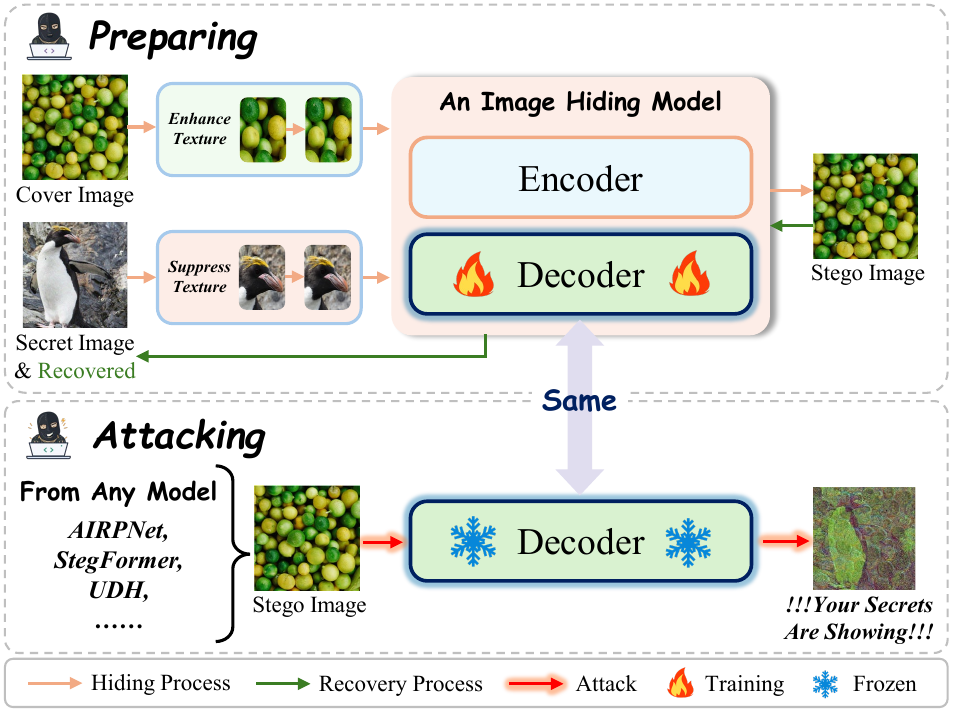}
\caption{\textbf{Overview of our ``secret showing'' attack}, which requires only two simple steps. \textbf{Preparing:} The attacker trains an image hiding model of their choice using its \emph{standard training procedure}. The attacker \emph{need not know or match the training datasets of the target models}. Based on our analysis, enhancing cover-image textures and suppressing secret-image textures can further improve secret extraction.
\textbf{Attacking:} The trained decoder is frozen and directly applied to stego images generated by \emph{unseen hiding models}, revealing the semantic content of the hidden images in a \emph{zero-shot} manner without target-specific training or fine-tuning. For the attacker, this decoder serves as a \emph{``secret extractor''}.}
\label{fig:intro2}
\end{center}
\end{figure}

How much can an attacker learn from what they extract? We can readily recognize the leaked content by eye, and our experiments show that machines can too. Using the extracted image as a query, a vision model can retrieve the original secret from a gallery, even for out-of-distribution secrets such as sketches, logos, and cartoons. The attack also works against multi-image hiding and against AIRPNet~\cite{AIRPNet}, a recent image restoration method that uses image steganography for privacy protection. Finally, our broader cross-model experiments reveal one more surprise. Extractors from \emph{all tested hiding models} appear to share this ``universal'' ability to varying degrees, revealing secret content even when the target model has a different architecture.

In summary, our main contributions are:
\textbf{(1)} We show that an independently trained decoder makes it surprisingly easy to reveal recognizable secret content from stego images produced by unseen hiding models, even in a zero-shot setting.
\textbf{(2)} We build an evaluation dataset from 11 hiding models and find that most of these models favor mid- and high-frequency embedding. More importantly, the decoder of every tested hiding model can act as a ``master key'' that reveals, to varying degrees, what other models hide.
\textbf{(3)} We show through image retrieval experiments that machines, like humans, can recognize the leaked content. We believe that, from now on, \emph{users can no longer rely on this form of covert communication to keep their secret images private}.
\section{Related Work}
\label{sec:related work}

\subsection{Deep Image Hiding}
Traditional image steganography embeds secret bits by modifying pixels or transform coefficients, using techniques such as least significant bit (LSB) substitution and content-adaptive distortion minimization~\cite{LSB,STC-arch,Digital-image-steganography}. Hiding an entire image, however, places much greater demands on embedding capacity. Baluja~\cite{baluja2019hiding} introduced a convolutional encoder-decoder framework that hides a full-sized color image within another image of the same size. Subsequent deep methods mainly fall into two families: autoencoders and INNs. Autoencoder-based methods learn separate hiding and revealing networks, while INN-based methods, such as HiNet~\cite{HiNet}, share parameters between the forward and inverse processes. Recent designs further incorporate Transformer and Mamba backbones and explore embedding in both spatial and wavelet domains~\cite{StegFormer,stegmamba}. \cref{tab:stego-quality} summarizes the representative models evaluated in this work.

\subsection{Privacy Protection Using Steganography}
\label{subsec:steganalysis and secret extraction attacks}
As deep image hiding or steganography continues to improve, researchers are increasingly using it to protect privacy in established vision tasks. For example, AIRPNet~\cite{AIRPNet} hides low-quality private images within cover images and performs restoration directly on the resulting stego images, allowing users to recover enhanced secret images afterward. StegaVAR~\cite{stegavar} hides action videos within ordinary cover videos and performs action recognition directly in the steganographic domain, without first extracting the secret videos. CloakDiff~\cite{clockdiff} uses diffusion-based adversarial editing to hinder privacy inference by vision-language models. It then embeds the original image into its adversarial counterpart through a hiding network, enabling subsequent recovery. In this work, we go beyond the steganalysis focus of prior security studies and reveal that deep image hiding is vulnerable to severe secret semantic leakage.

\begin{table}[!h]
  \caption{\textbf{Overview of the representative hiding models.} The two rightmost columns report the PSNR (dB) of stego images.}
  \label{tab:stego-quality}
  \centering
  \scriptsize
  \setlength{\tabcolsep}{0.8pt}
  \renewcommand{\arraystretch}{1.1}
  \setlength{\dashlinedash}{3pt}
  \setlength{\dashlinegap}{3pt}
  \newcommand{\overviewhead}[1]{%
    \multirow[c]{2}{*}[-\dimexpr(\aboverulesep+\belowrulesep+\cmidrulewidth)/2\relax]{#1}%
  }

  \begin{tabularx}{\columnwidth}{
    @{}clc>{\centering\arraybackslash}X*{2}{c}@{}
  }
    \toprule
    \overviewhead{Family}
      & \overviewhead{Method}
      & \overviewhead{Backbone}
      & \overviewhead{Domain}
      & COCO
      & ImageNet \\
        \cmidrule(r{1pt}){5-5}
        \cmidrule(l{1pt}){6-6}
      & & & & PSNR$\uparrow$ & PSNR$\uparrow$ \\
    \midrule
    \multirow{5}{*}{INN}
      & HiNet~\cite{HiNet} & CNN & Wavelet
      & 46.02 & 46.04 \\
      & LiDiNet~\cite{lidinet} & CNN & Wavelet
      & 46.89 & 47.13 \\
      & StegFlow~\cite{stegflow} & CNN & Wavelet
      & 47.78 & 47.94 \\
      & StegMamba~\cite{stegmamba} & \shortstack[c]{CNN + Mamba} & Wavelet
      & 47.01 & 47.15 \\
      & EFCA-DIH~\cite{EFCA} & CNN & Wavelet
      & 45.28 & 45.38 \\
    \hdashline
    \multirow{6}{*}{\shortstack[c]{Auto\\encoder}}
      & Baluja~\cite{baluja2019hiding} & CNN & Spatial
      & 37.02 & 36.80 \\
      & StegFormer~\cite{StegFormer} & Transformer & Spatial
      & 50.80 & 50.94 \\
      & Ectformer~\cite{ectformer} & CNN + Transformer & EMD
      & 45.07 & 45.10 \\
      & StegTransX~\cite{StegTransX} & \shortstack[c]{CNN + Transformer} & Spatial
      & 46.71 & 47.04 \\
      & UDH~\cite{udh} & CNN & Spatial
      & 39.63 & 39.60 \\
      & URSMamba~\cite{ursmamba} & Mamba & Spatial + Wavelet
      & 42.37 & 42.26 \\
    \bottomrule
  \end{tabularx}
  \par\vspace{2pt}
  \begin{minipage}{\columnwidth}
  \scriptsize\raggedright
  ``Wavelet'' denotes the discrete wavelet transform domain, which decomposes images into frequency sub-bands. ``EMD'' denotes empirical mode decomposition.
\end{minipage}
\end{table}
\section{Uncovering Cross-Model Secret Leakage}
\label{sec:motivation}
In this section, we start with an interesting finding, build an evaluation dataset to take a first look at why it happens, and introduce simple yet effective ways to strengthen the attack.





\begin{figure*}[!t]
\begin{center}
\includegraphics[width=15cm]{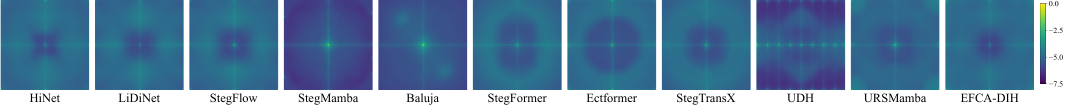}
\caption{\textbf{Frequency analysis of embedding residuals.} Each panel shows the normalized mean power spectrum of the stego-cover residuals for one model, computed from 7,550 image pairs drawn from the DIV2K, COCO, and ImageNet test datasets.}
\label{fig:freq_analysis}
\end{center}
\vspace{-1em}
\end{figure*}

\subsection{From Detection to Secret Extraction}
\label{subsec:vulnerability observation}

In deep image hiding or image-in-image steganography, a cover image and a secret image are passed through an encoder to produce a stego image. The secret image can then be recovered from the stego image using the corresponding decoder. In general, an attacker does not know which encoder generated the stego image they obtain (and, of course, does not know the corresponding decoder either). They therefore have to rely on rich-model-based~\cite{SRM,maxSRMd2} or DNN-based~\cite{SRNet,Zhu-Net} steganalyzers to determine whether an image is a stego image. This is a binary detection task. Once a stego image is detected, the covert communication is exposed, and the attacker can refuse to forward it. However, they still cannot tell what is hidden inside.

\textbf{Shockingly, now they can.} We find that an attacker can obtain a ``universal'' secret extractor with surprisingly little effort. As shown in~\cref{fig:intro2}, they can obtain a decoder by independently training an image hiding model of their choice using its standard training procedure. During this process, they need neither to refer to other models nor to learn from the stego images those models generate. Here, we choose HiNet~\cite{HiNet}, a high-performing open-source hiding model, as a representative example. The resulting decoder can then be used to extract secrets from stego images produced by any of the tested models, as shown in the ``\textbf{Attack}'' row of~\cref{fig:main-visual}. We can clearly see that the secret images' key texture patterns are exposed, although the extracted results still contain some interference from the stego images. As humans, we can already tell what the secret images depict from these leaked cues alone. In fact, our later experiments show that vision networks can do this too.

\subsection{Can We Make the Attack Even Stronger?}
\label{subsec:threat model}

Given this finding, we cannot help wondering what causes it (see ~\cref{subsubsec:resnet-attack} for further discussion). But an attacker would surely want to find a \emph{simple and effective} way to strengthen this attack and make the leaked content clearer. To this end, we construct an evaluation dataset using 11 image hiding models with publicly available or official implementations. They belong to different families (INNs and autoencoders), use different backbones (CNN, Transformer, and Mamba), and embed information in different domains (including spatial and frequency). We perform a frequency analysis of the 83,050 stego images generated by these models. Specifically, we first compute the residual between each stego image and its corresponding cover image:

\begin{equation}
\Delta_{res} = I_{\mathit{stego}} - I_{\mathit{cover}}.
\label{eq:embedding_residual}
\end{equation}
where $\Delta_{\mathit{res}}$ denotes the embedding residual. Adapting the frequency analysis of Wang et al.~\cite{CNN-Generated}, we visualize the normalized power spectra of the $\Delta_{res}$ in ~\cref{fig:freq_analysis}. It is easy to see that most of these methods show a “preference” for embedding in the mid- and high-frequency components. Based on this finding, a simple idea comes to mind: use data augmentation to reinforce this preference during HiNet~\cite{HiNet} training. Specifically, we enhance textures in the cover images while suppressing textures in the secret images, and denote the resulting variant by \textbf{ours+}. In addition, we can increase the weight of HiNet’s~\cite{HiNet} low-frequency loss (referred to as \textbf{ours++}). Both modifications encourage the network to focus more on the cover images’ high-frequency components when hiding information. Our later experiments show that both approaches effectively improve the performance and generalization of cross-model secret extraction. Although we discovered this preference by analyzing 11 models, most of them exhibit it, so an attacker may well observe it using only stego-cover pairs from their own trained model.

\section{Building Transferable Secret Extractors}
\label{sec:secret-extraction}
We now discuss the experimental setup for attacking unseen hiding models and evaluate our secret extractors.

\subsection{Experimental Setup}
\label{subsec:experimental-setup}
To facilitate comparison, we start with one representative model and evaluate extractors from all tested models in the final discussion. Here, we still choose HiNet~\cite{HiNet} to train our “secret extractor”. Training follows the official settings. Details of the settings are provided in the supplemental material. For the augmentation (\textbf{ours+}), we enhance textures in the cover images while suppressing textures in the secret images during training. For the second augmentation (\textbf{ours++}), we build on the first and double the weight $\lambda_f$ of the low-frequency loss. Unless otherwise specified, we conduct experiments on COCO~\cite{COCO} and ImageNet~\cite{ImageNet}, using 5,000 and 10,000 randomly selected images, respectively.

\subsection{Cross-Model Secret Extraction}
\label{subsec:cross-model-extraction}

\subsubsection*{Qualitative results.}
\begin{figure*}[!t]
\begin{center}
\includegraphics[width=16.2cm]{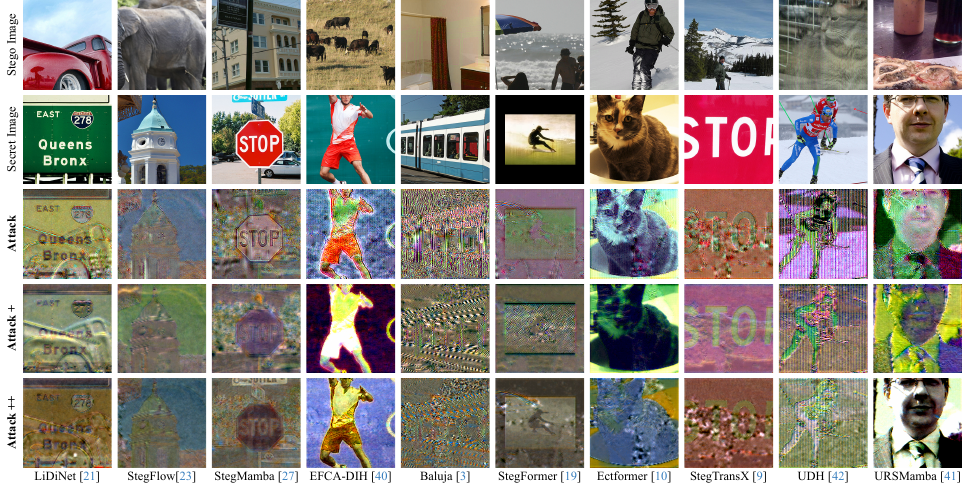}
\caption{\textbf{Visual results.} In each column, the first row shows a stego image generated by the indicated target model, and the second shows the original secret image hidden in it. The remaining rows show the content recovered from that same stego image by our three attacks.}
\label{fig:main-visual}
\end{center}
\vspace{-1em}
\end{figure*}

We first show the visual results, as they provide the most intuitive demonstration of this type of attack. \cref{fig:main-visual} shows the results of attacking stego images generated by other models using only HiNet's decoder~\cite{HiNet}. As can be seen in the ``\textbf{Attack}'' row, the extracted semantic content closely matches that of the original secret images (\emph{e.g.}, a parachute, the word ``STOP'', a cat, and a human face). With the proposed augmentations, ``\textbf{Attack+}'' and ``\textbf{Attack++}'' further improve the extraction results to varying degrees (\emph{e.g.}, the action depicted in the EFCA-DIH~\cite{EFCA} column becomes clearer). Just imagine that an attacker knows how to carry out this attack. Judging from these visual results, using deep image hiding for covert communication may now be little different from sending the secret image directly, in terms of what humans can recognize. We believe this should raise serious concern: \emph{hiding an image no longer guarantees that its content remains private}. By revealing the hidden content itself, this attack goes beyond the binary detection performed by conventional steganalyzers.

\begin{table*}[!t]
\caption{\textbf{Cross-model secret retrieval.} We compare three attacks against 10 unseen models on two datasets. Retrieval uses cosine similarity between CLS token embeddings from the final layer of DINOv3 ViT-L/16~\cite{dinov3}. Blue shading marks the best retrieval setting for each target.}
  \label{tab:secret-extraction-retrieval}
  \definecolor{attackblue}{HTML}{E6F0FA}
  \centering
  \scriptsize
  \setlength{\tabcolsep}{1pt}
  \renewcommand{\arraystretch}{1}
  \setlength{\dashlinedash}{3pt}
  \setlength{\dashlinegap}{3pt}

  \begin{tabularx}{\textwidth}{
    @{}ccl
    *{4}{>{\centering\arraybackslash}X}
    *{2}{>{\centering\arraybackslash}X}
    *{4}{>{\centering\arraybackslash}X}
    *{2}{>{\centering\arraybackslash}X}@{}
  }
    \toprule

    \multirow{4}{*}{\shortstack[c]{Attack}}
      & \multirow{4}{*}{Family}
      & \multirow{4}{*}{Target model}
      & \multicolumn{6}{c}{COCO}
      & \multicolumn{6}{c}{ImageNet} \\
    \cmidrule(lr){4-9}
    \cmidrule(l){10-15}

      & &
      & \multicolumn{4}{c}{Secret retrieval (\%)}
      & \multicolumn{2}{c}{Secret quality}
      & \multicolumn{4}{c}{Secret retrieval (\%)}
      & \multicolumn{2}{c}{Secret quality} \\
    \cmidrule(lr){4-7}
    \cmidrule(lr){8-9}
    \cmidrule(lr){10-13}
    \cmidrule(l){14-15}

      & &
      & R@1$\uparrow$ & R@10$\uparrow$ & R@20$\uparrow$ & R@50$\uparrow$
      & PSNR$\uparrow$ & SSIM$\uparrow$
      & R@1$\uparrow$ & R@10$\uparrow$ & R@20$\uparrow$ & R@50$\uparrow$
      & PSNR$\uparrow$ & SSIM$\uparrow$ \\
    \midrule

    \multirow{10}{*}{\textbf{ours}} & \multirow{4}{*}{INN} & LiDiNet~\cite{lidinet}
      & \textcolor{black!60}{8.20} & \textcolor{black!60}{16.16} & \textcolor{black!60}{19.44} & \textcolor{black!60}{25.32} & \textcolor{black!60}{10.23} & \textcolor{black!60}{0.068}
      & \textcolor{black!60}{3.88} & \textcolor{black!60}{8.04} & \textcolor{black!60}{10.02} & \textcolor{black!60}{12.88} & \textcolor{black!60}{11.64} & \textcolor{black!60}{0.136} \\
     &  & StegFlow~\cite{stegflow}
      & \textcolor{black!60}{3.44} & \textcolor{black!60}{7.72} & \textcolor{black!60}{10.44} & \textcolor{black!60}{14.00} & \textcolor{black!60}{12.31} & \textcolor{black!60}{0.116}
      & \textcolor{black!60}{3.20} & \textcolor{black!60}{7.58} & \textcolor{black!60}{9.56} & \textcolor{black!60}{13.16} & \textcolor{black!60}{12.42} & \textcolor{black!60}{0.116} \\
     &  & StegMamba~\cite{stegmamba}
      & \textcolor{black!60}{6.72} & \textcolor{black!60}{14.92} & \textcolor{black!60}{18.12} & \textcolor{black!60}{23.60} & \textcolor{black!60}{11.41} & \textcolor{black!60}{0.128}
      & \textcolor{black!60}{7.34} & \textcolor{black!60}{14.74} & \textcolor{black!60}{17.44} & \textcolor{black!60}{21.78} & \textcolor{black!60}{11.51} & \textcolor{black!60}{0.125} \\
     &  & EFCA-DIH~\cite{EFCA}
      & \textcolor{black!60}{30.96} & \textcolor{black!60}{50.84} & \textcolor{black!60}{56.64} & \textcolor{black!60}{63.88} & \textcolor{black!60}{11.84} & \textcolor{black!60}{0.073}
      & \textcolor{black!60}{30.62} & \textcolor{black!60}{48.88} & \textcolor{black!60}{53.44} & \textcolor{black!60}{59.30} & \textcolor{black!60}{11.91} & \textcolor{black!60}{0.080} \\
    \cdashline{2-15}
     & \multirow{6}{*}{\shortstack[c]{Auto\\encoder}} & Baluja~\cite{baluja2019hiding}
      & \textcolor{black!60}{0.60} & \textcolor{black!60}{2.92} & \textcolor{black!60}{4.96} & \textcolor{black!60}{8.04} & \textcolor{black!60}{9.29} & \textcolor{black!60}{0.019}
      & \textcolor{black!60}{1.48} & \textcolor{black!60}{4.28} & \textcolor{black!60}{5.72} & \textcolor{black!60}{8.64} & \textcolor{black!60}{9.80} & \textcolor{black!60}{0.026} \\
     &  & StegFormer~\cite{StegFormer}
      & \textcolor{black!60}{0.72} & \textcolor{black!60}{2.56} & \textcolor{black!60}{3.44} & \textcolor{black!60}{5.52} & \textcolor{black!60}{12.04} & \textcolor{black!60}{0.155}
      & \textcolor{black!60}{0.62} & \textcolor{black!60}{1.92} & \textcolor{black!60}{2.64} & \textcolor{black!60}{4.20} & \textcolor{black!60}{12.14} & \textcolor{black!60}{0.155} \\
     &  & Ectformer~\cite{ectformer}
      & \textcolor{black!60}{64.56} & \textcolor{black!60}{80.68} & \textcolor{black!60}{83.84} & \textcolor{black!60}{87.12} & \textcolor{black!60}{14.54} & \textcolor{black!60}{0.172}
      & \textcolor{black!60}{66.50} & \textcolor{black!60}{80.26} & \textcolor{black!60}{82.74} & \textcolor{black!60}{85.92} & \textcolor{black!60}{14.61} & \textcolor{black!60}{0.173} \\
     &  & StegTransX~\cite{StegTransX}
      & \textcolor{black!60}{5.08} & \textcolor{black!60}{11.48} & \textcolor{black!60}{14.16} & \textcolor{black!60}{18.64} & \textcolor{black!60}{12.24} & \textcolor{black!60}{0.138}
      & \textcolor{black!60}{6.24} & \textcolor{black!60}{12.28} & \textcolor{black!60}{14.94} & \textcolor{black!60}{18.18} & \textcolor{black!60}{12.32} & \textcolor{black!60}{0.138} \\
     &  & \cellcolor{attackblue}UDH~\cite{udh}
      & \cellcolor{attackblue}\textcolor{black}{23.32} & \cellcolor{attackblue}\textcolor{black}{42.76} & \cellcolor{attackblue}\textcolor{black}{48.92} & \cellcolor{attackblue}\textcolor{black}{56.44} & \cellcolor{attackblue}\textcolor{black!60}{9.05} & \cellcolor{attackblue}\textcolor{black!60}{0.042}
      & \cellcolor{attackblue}\textcolor{black}{25.04} & \cellcolor{attackblue}\textcolor{black}{41.76} & \cellcolor{attackblue}\textcolor{black}{46.86} & \cellcolor{attackblue}\textcolor{black}{53.58} & \cellcolor{attackblue}\textcolor{black!60}{8.97} & \cellcolor{attackblue}\textcolor{black!60}{0.039} \\
     &  & URSMamba~\cite{ursmamba}
      & \textcolor{black!60}{35.20} & \textcolor{black!60}{55.12} & \textcolor{black!60}{61.12} & \textcolor{black!60}{68.20} & \textcolor{black!60}{11.05} & \textcolor{black!60}{0.166}
      & \textcolor{black!60}{37.28} & \textcolor{black!60}{55.78} & \textcolor{black!60}{60.18} & \textcolor{black!60}{66.02} & \textcolor{black!60}{11.23} & \textcolor{black!60}{0.163} \\
    \cmidrule{1-15}

    \multirow{10}{*}{\textbf{\shortstack[c]{ours\\+}}} & \multirow{4}{*}{INN} & \cellcolor{attackblue}LiDiNet~\cite{lidinet}
      & \cellcolor{attackblue}\textcolor{black}{9.64} & \cellcolor{attackblue}\textcolor{black}{19.64} & \cellcolor{attackblue}\textcolor{black}{23.36} & \cellcolor{attackblue}\textcolor{black}{28.92} & \cellcolor{attackblue}\textcolor{black!60}{12.29} & \cellcolor{attackblue}\textcolor{black!60}{0.250}
      & \cellcolor{attackblue}\textcolor{black}{10.68} & \cellcolor{attackblue}\textcolor{black}{20.68} & \cellcolor{attackblue}\textcolor{black}{23.54} & \cellcolor{attackblue}\textcolor{black}{28.94} & \cellcolor{attackblue}\textcolor{black!60}{10.43} & \cellcolor{attackblue}\textcolor{black!60}{0.154} \\
     &  & \cellcolor{attackblue}StegFlow~\cite{stegflow}
      & \cellcolor{attackblue}\textcolor{black}{25.64} & \cellcolor{attackblue}\textcolor{black}{41.36} & \cellcolor{attackblue}\textcolor{black}{46.36} & \cellcolor{attackblue}\textcolor{black}{52.76} & \cellcolor{attackblue}\textcolor{black!60}{14.92} & \cellcolor{attackblue}\textcolor{black!60}{0.308}
      & \cellcolor{attackblue}\textcolor{black}{27.48} & \cellcolor{attackblue}\textcolor{black}{42.20} & \cellcolor{attackblue}\textcolor{black}{45.12} & \cellcolor{attackblue}\textcolor{black}{50.36} & \cellcolor{attackblue}\textcolor{black!60}{15.04} & \cellcolor{attackblue}\textcolor{black!60}{0.306} \\
     &  & StegMamba~\cite{stegmamba}
      & \textcolor{black!60}{3.92} & \textcolor{black!60}{10.32} & \textcolor{black!60}{12.72} & \textcolor{black!60}{17.52} & \textcolor{black!60}{12.53} & \textcolor{black!60}{0.265}
      & \textcolor{black!60}{3.90} & \textcolor{black!60}{9.30} & \textcolor{black!60}{11.44} & \textcolor{black!60}{15.40} & \textcolor{black!60}{12.69} & \textcolor{black!60}{0.264} \\
     &  & EFCA-DIH~\cite{EFCA}
      & \textcolor{black!60}{57.92} & \textcolor{black!60}{76.72} & \textcolor{black!60}{80.68} & \textcolor{black!60}{85.36} & \textcolor{black!60}{15.23} & \textcolor{black!60}{0.303}
      & \textcolor{black!60}{57.92} & \textcolor{black!60}{75.88} & \textcolor{black!60}{78.84} & \textcolor{black!60}{82.56} & \textcolor{black!60}{15.20} & \textcolor{black!60}{0.308} \\
    \cdashline{2-15}
     & \multirow{6}{*}{\shortstack[c]{Auto\\encoder}} & \cellcolor{attackblue}Baluja~\cite{baluja2019hiding}
      & \cellcolor{attackblue}\textcolor{black}{2.12} & \cellcolor{attackblue}\textcolor{black}{6.40} & \cellcolor{attackblue}\textcolor{black}{9.12} & \cellcolor{attackblue}\textcolor{black}{13.52} & \cellcolor{attackblue}\textcolor{black!60}{11.86} & \cellcolor{attackblue}\textcolor{black!60}{0.050}
      & \cellcolor{attackblue}\textcolor{black}{1.96} & \cellcolor{attackblue}\textcolor{black}{6.46} & \cellcolor{attackblue}\textcolor{black}{8.06} & \cellcolor{attackblue}\textcolor{black}{11.10} & \cellcolor{attackblue}\textcolor{black!60}{12.01} & \cellcolor{attackblue}\textcolor{black!60}{0.052} \\
     &  & StegFormer~\cite{StegFormer}
      & \textcolor{black!60}{1.44} & \textcolor{black!60}{3.64} & \textcolor{black!60}{5.08} & \textcolor{black!60}{7.84} & \textcolor{black!60}{13.60} & \textcolor{black!60}{0.203}
      & \textcolor{black!60}{1.06} & \textcolor{black!60}{3.16} & \textcolor{black!60}{3.96} & \textcolor{black!60}{6.06} & \textcolor{black!60}{13.45} & \textcolor{black!60}{0.215} \\
     &  & \cellcolor{attackblue}Ectformer~\cite{ectformer}
      & \cellcolor{attackblue}\textcolor{black}{74.48} & \cellcolor{attackblue}\textcolor{black}{86.08} & \cellcolor{attackblue}\textcolor{black}{88.84} & \cellcolor{attackblue}\textcolor{black}{91.32} & \cellcolor{attackblue}\textcolor{black!60}{17.01} & \cellcolor{attackblue}\textcolor{black!60}{0.476}
      & \cellcolor{attackblue}\textcolor{black}{77.14} & \cellcolor{attackblue}\textcolor{black}{87.04} & \cellcolor{attackblue}\textcolor{black}{89.02} & \cellcolor{attackblue}\textcolor{black}{91.42} & \cellcolor{attackblue}\textcolor{black!60}{17.14} & \cellcolor{attackblue}\textcolor{black!60}{0.479} \\
     &  & \cellcolor{attackblue}StegTransX~\cite{StegTransX}
      & \cellcolor{attackblue}\textcolor{black}{23.60} & \cellcolor{attackblue}\textcolor{black}{37.80} & \cellcolor{attackblue}\textcolor{black}{42.40} & \cellcolor{attackblue}\textcolor{black}{48.64} & \cellcolor{attackblue}\textcolor{black!60}{15.60} & \cellcolor{attackblue}\textcolor{black!60}{0.348}
      & \cellcolor{attackblue}\textcolor{black}{26.52} & \cellcolor{attackblue}\textcolor{black}{39.02} & \cellcolor{attackblue}\textcolor{black}{42.48} & \cellcolor{attackblue}\textcolor{black}{47.16} & \cellcolor{attackblue}\textcolor{black!60}{15.72} & \cellcolor{attackblue}\textcolor{black!60}{0.346} \\
     &  & UDH~\cite{udh}
      & \textcolor{black!60}{5.00} & \textcolor{black!60}{12.64} & \textcolor{black!60}{15.92} & \textcolor{black!60}{20.76} & \textcolor{black!60}{11.67} & \textcolor{black!60}{0.070}
      & \textcolor{black!60}{6.14} & \textcolor{black!60}{12.22} & \textcolor{black!60}{14.92} & \textcolor{black!60}{18.84} & \textcolor{black!60}{11.77} & \textcolor{black!60}{0.070} \\
     &  & URSMamba~\cite{ursmamba}
      & \textcolor{black!60}{23.24} & \textcolor{black!60}{42.88} & \textcolor{black!60}{48.88} & \textcolor{black!60}{57.12} & \textcolor{black!60}{13.41} & \textcolor{black!60}{0.161}
      & \textcolor{black!60}{24.64} & \textcolor{black!60}{42.78} & \textcolor{black!60}{47.48} & \textcolor{black!60}{54.36} & \textcolor{black!60}{13.46} & \textcolor{black!60}{0.160} \\
    \cmidrule{1-15}
     \multirow{10}{*}{\textbf{\shortstack[c]{ours\\++}}} & \multirow{4}{*}{INN} & LiDiNet~\cite{lidinet}
      & \textcolor{black!60}{6.52} & \textcolor{black!60}{14.52} & \textcolor{black!60}{17.64} & \textcolor{black!60}{23.48} & \textcolor{black!60}{11.05} & \textcolor{black!60}{0.159}
      & \textcolor{black!60}{7.26} & \textcolor{black!60}{14.92} & \textcolor{black!60}{17.84} & \textcolor{black!60}{22.28} & \textcolor{black!60}{11.17} & \textcolor{black!60}{0.154} \\
     &  & StegFlow~\cite{stegflow}
      & \textcolor{black!60}{4.44} & \textcolor{black!60}{10.48} & \textcolor{black!60}{13.04} & \textcolor{black!60}{17.52} & \textcolor{black!60}{12.42} & \textcolor{black!60}{0.253}
      & \textcolor{black!60}{5.08} & \textcolor{black!60}{11.42} & \textcolor{black!60}{13.72} & \textcolor{black!60}{16.78} & \textcolor{black!60}{12.53} & \textcolor{black!60}{0.249} \\
     &  & \cellcolor{attackblue}StegMamba~\cite{stegmamba}
      & \cellcolor{attackblue}\textcolor{black}{29.88} & \cellcolor{attackblue}\textcolor{black}{48.16} & \cellcolor{attackblue}\textcolor{black}{53.44} & \cellcolor{attackblue}\textcolor{black}{59.64} & \cellcolor{attackblue}\textcolor{black!60}{12.25} & \cellcolor{attackblue}\textcolor{black!60}{0.184}
      & \cellcolor{attackblue}\textcolor{black}{28.98} & \cellcolor{attackblue}\textcolor{black}{44.66} & \cellcolor{attackblue}\textcolor{black}{48.96} & \cellcolor{attackblue}\textcolor{black}{54.58} & \cellcolor{attackblue}\textcolor{black!60}{12.34} & \cellcolor{attackblue}\textcolor{black!60}{0.177} \\
     &  & \cellcolor{attackblue}EFCA-DIH~\cite{EFCA}
      & \cellcolor{attackblue}\textcolor{black}{70.04} & \cellcolor{attackblue}\textcolor{black}{85.20} & \cellcolor{attackblue}\textcolor{black}{88.32} & \cellcolor{attackblue}\textcolor{black}{91.52} & \cellcolor{attackblue}\textcolor{black!60}{17.16} & \cellcolor{attackblue}\textcolor{black!60}{0.464}
      & \cellcolor{attackblue}\textcolor{black}{72.56} & \cellcolor{attackblue}\textcolor{black}{85.54} & \cellcolor{attackblue}\textcolor{black}{88.04} & \cellcolor{attackblue}\textcolor{black}{90.68} & \cellcolor{attackblue}\textcolor{black!60}{17.24} & \cellcolor{attackblue}\textcolor{black!60}{0.470} \\
    \cdashline{2-15}
     & \multirow{6}{*}{\shortstack[c]{Auto\\encoder}} & Baluja~\cite{baluja2019hiding}
      & \textcolor{black!60}{0.32} & \textcolor{black!60}{1.96} & \textcolor{black!60}{3.20} & \textcolor{black!60}{6.36} & \textcolor{black!60}{12.22} & \textcolor{black!60}{0.077}
      & \textcolor{black!60}{0.42} & \textcolor{black!60}{1.98} & \textcolor{black!60}{3.00} & \textcolor{black!60}{4.82} & \textcolor{black!60}{12.23} & \textcolor{black!60}{0.077} \\
     &  & \cellcolor{attackblue}StegFormer~\cite{StegFormer}
      & \cellcolor{attackblue}\textcolor{black}{13.80} & \cellcolor{attackblue}\textcolor{black}{22.88} & \cellcolor{attackblue}\textcolor{black}{25.76} & \cellcolor{attackblue}\textcolor{black}{30.88} & \cellcolor{attackblue}\textcolor{black!60}{14.37} & \cellcolor{attackblue}\textcolor{black!60}{0.331}
      & \cellcolor{attackblue}\textcolor{black}{14.68} & \cellcolor{attackblue}\textcolor{black}{22.46} & \cellcolor{attackblue}\textcolor{black}{25.24} & \cellcolor{attackblue}\textcolor{black}{28.70} & \cellcolor{attackblue}\textcolor{black!60}{14.40} & \cellcolor{attackblue}\textcolor{black!60}{0.328} \\
     &  & Ectformer~\cite{ectformer}
      & \textcolor{black!60}{27.84} & \textcolor{black!60}{45.96} & \textcolor{black!60}{51.28} & \textcolor{black!60}{58.52} & \textcolor{black!60}{12.99} & \textcolor{black!60}{0.176}
      & \textcolor{black!60}{27.94} & \textcolor{black!60}{43.94} & \textcolor{black!60}{47.80} & \textcolor{black!60}{53.64} & \textcolor{black!60}{13.04} & \textcolor{black!60}{0.171} \\
     &  & StegTransX~\cite{StegTransX}
      & \textcolor{black!60}{1.80} & \textcolor{black!60}{4.60} & \textcolor{black!60}{6.04} & \textcolor{black!60}{9.08} & \textcolor{black!60}{13.20} & \textcolor{black!60}{0.280}
      & \textcolor{black!60}{2.12} & \textcolor{black!60}{5.56} & \textcolor{black!60}{6.98} & \textcolor{black!60}{9.58} & \textcolor{black!60}{13.26} & \textcolor{black!60}{0.275} \\
     &  & UDH~\cite{udh}
      & \textcolor{black!60}{4.72} & \textcolor{black!60}{12.48} & \textcolor{black!60}{15.32} & \textcolor{black!60}{21.40} & \textcolor{black!60}{11.90} & \textcolor{black!60}{0.066}
      & \textcolor{black!60}{6.74} & \textcolor{black!60}{12.90} & \textcolor{black!60}{15.42} & \textcolor{black!60}{20.16} & \textcolor{black!60}{11.88} & \textcolor{black!60}{0.064} \\
     &  & \cellcolor{attackblue}URSMamba~\cite{ursmamba}
      & \cellcolor{attackblue}\textcolor{black}{96.60} & \cellcolor{attackblue}\textcolor{black}{99.40} & \cellcolor{attackblue}\textcolor{black}{99.68} & \cellcolor{attackblue}\textcolor{black}{99.80} & \cellcolor{attackblue}\textcolor{black!60}{17.74} & \cellcolor{attackblue}\textcolor{black!60}{0.556}
      & \cellcolor{attackblue}\textcolor{black}{97.22} & \cellcolor{attackblue}\textcolor{black}{99.08} & \cellcolor{attackblue}\textcolor{black}{99.36} & \cellcolor{attackblue}\textcolor{black}{99.68} & \cellcolor{attackblue}\textcolor{black!60}{17.84} & \cellcolor{attackblue}\textcolor{black!60}{0.561} \\

    \bottomrule
  \end{tabularx}
\vspace{-1em}
\end{table*}

\subsubsection*{Quantitative results.} Although the attack results in \cref{fig:main-visual} are striking, quantifying how much secret content is revealed is not straightforward, as the extracted images also contain interference from the stego images (more precisely, from the cover images). Alongside PSNR~\cite{PSNR} and SSIM~\cite{SSIM}, we therefore introduce a more practical evaluation inspired by image retrieval~\cite{Recall-1,Recall-2}. Specifically, an attacker feeds an extracted image into a vision model (we use DINOv3 ViT-L/16~\cite{dinov3}, a self-supervised vision foundation model) to retrieve the corresponding original secret image from a large candidate gallery. We report Recall@$K$ (R@$K$), defined here as the percentage of queries for which the original secret image appears among the top-$K$ retrieved candidates. Because the extracted images retain traces of their covers, we restrict each candidate gallery to the full set of original secret images used in that evaluation.

As shown in~\cref{tab:secret-extraction-retrieval}, PSNR and SSIM do not fully reflect whether the core semantics of a secret image have been leaked. PSNR is based on pixelwise errors, while SSIM averages local similarity scores across the image. Both are therefore affected by interference from the stego or cover image. In contrast, ``secret retrieval'' using the extracted content provides a more direct assessment of such leakage. Specifically, even without augmentation (\textbf{ours}), the attack is already effective against stego images generated by most of the evaluated models on COCO. For UDH~\cite{udh}, R@1, R@10, R@20, and R@50 reach 23.32\%, 42.76\%, 48.92\%, and 56.44\%, respectively. This means that the original secret image ranks first for nearly one in four queries and appears among the top 50 candidates for more than half of them. With augmentation (\textbf{ours+}), attacks on LiDiNet~\cite{lidinet}, StegFlow~\cite{stegflow}, and Ectformer~\cite{ectformer} improve more noticeably, reaching R@10 scores of 19.64\%, 41.36\%, and 86.08\%, respectively. Here, a successful retrieval places the original secret image among just ten candidates from the large gallery. For Ectformer~\cite{ectformer}, this happens in about 86\% of cases. The second augmentation (\textbf{ours++}) further improves attacks on StegMamba~\cite{stegmamba}, StegFormer~\cite{StegFormer}, and URSMamba~\cite{ursmamba}, with even their R@1 scores reaching 29.88\%, 13.80\%, and 96.60\%, respectively. An additional control experiment shows that using stego images directly as queries fails to retrieve the original secrets (see the supplementary material).

That being said, we must also acknowledge that the gains from the two augmentations are not uniform across models, as their preferences for embedding differ in strength. What we want to emphasize here, however, is that these augmentations are easy for an attacker to implement and provide simple yet effective ways to strengthen the attack.

\subsection{Generalization Across Secret Domains}
\label{subsec:across-secret-domains}

\begin{figure}[!t]
\begin{center}
\includegraphics[width=7.5cm]{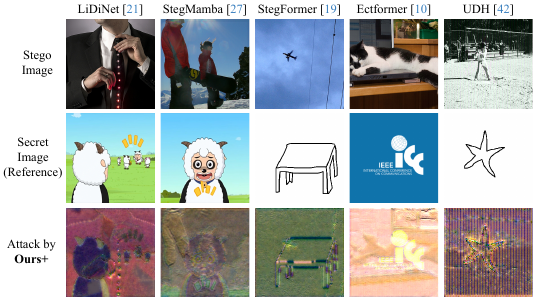}
\caption{\textbf{Visual results of our attack on out-of-distribution secrets.} We use the attack with augmentation (\textbf{ours+}).}
\label{fig:secret-domains}
\end{center}
\vspace{-0.3em}
\end{figure}

\begin{table}[!t]
   \caption{\textbf{Secret retrieval on sketches, logos, and cartoons.} For most target models, both R@1 exceed 50\% on sketches and logos.}
  \label{tab:sketches-logos-cartoons}
  \centering
  \footnotesize
  \setlength{\tabcolsep}{2pt}
  \renewcommand{\arraystretch}{1}

  \begin{tabularx}{\columnwidth}{
    @{}l*{6}{>{\centering\arraybackslash}X}@{}
  }
    \toprule
    \multirow[c]{2}{*}[-\dimexpr(\aboverulesep+\belowrulesep+\cmidrulewidth)/2\relax]{Target Model}
      & \multicolumn{2}{c}{Sketch}
      & \multicolumn{2}{c}{Logo}
      & \multicolumn{2}{c}{Cartoon} \\
    \cmidrule(lr){2-3}
    \cmidrule(lr){4-5}
    \cmidrule(l){6-7}
      & R@1$\uparrow$ & R@10$\uparrow$
      & R@1$\uparrow$ & R@10$\uparrow$
      & R@1$\uparrow$ & R@10$\uparrow$ \\
    \midrule
    HiNet~\cite{HiNet} & 68.50 & 86.30 & 82.00 & 93.90 & 53.70 & 73.20 \\
    LiDiNet~\cite{lidinet} & 55.70 & 73.80 & 50.80 & 73.40 & 10.40 & 22.80 \\
    StegFlow~\cite{stegflow} & 51.80 & 72.60 & 61.00 & 81.30 & 25.40 & 44.80 \\
    StegMamba~\cite{stegmamba} & 44.90 & 69.30 & 38.40 & 60.10 & 6.60 & 16.90 \\
    EFCA-DIH~\cite{EFCA} & 67.20 & 87.80 & 85.40 & 95.50 & 58.30 & 79.50 \\
    \cdashline{1-7}
    Baluja~\cite{baluja2019hiding} & 13.50 & 31.30 & 22.40 & 44.50 & 1.30 & 5.10 \\
    StegFormer~\cite{StegFormer} & 24.60 & 43.90 & 14.90 & 27.10 & 1.30 & 4.20 \\
    Ectformer~\cite{ectformer} & 86.40 & 94.80 & 81.00 & 90.80 & 66.90 & 85.30 \\
    StegTransX~\cite{StegTransX} & 54.70 & 71.20 & 52.70 & 72.80 & 18.90 & 37.10 \\
    UDH~\cite{udh} & 18.40 & 33.80 & 21.70 & 39.80 & 5.20 & 13.90 \\
    URSMamba~\cite{ursmamba} & 42.20 & 68.60 & 69.80 & 88.40 & 25.60 & 45.10 \\
    \bottomrule
  \end{tabularx}
\vspace{-0.5em}
\end{table}

Legitimate users may hide many different types of secret images for covert communication. This naturally raises a question for an attacker: \emph{does the attack still work when the secret images are out of distribution?} To find out, we select datasets of sketches~\cite{sketchy}, logos~\cite{logo}, and cartoons~\cite{cartoon} as secret images and hide them in cover images from COCO.

As shown in~\cref{fig:secret-domains}, the secret content extracted across the five representative target models remains clearly recognizable. \cref{tab:sketches-logos-cartoons} summarizes the secret retrieval results for all 11 target models. For sketches and logos, R@1 exceeds 50\% for six and seven models, respectively. This means that the original secret image is already the top-ranked result for more than half of the queries for these models. R@10 exceeds 50\% for eight models in each of these two domains, meaning that the original secret image appears among just ten retrieved candidates in more than half of the cases. For cartoons, HiNet~\cite{HiNet}, Ectformer~\cite{ectformer}, and EFCA-DIH~\cite{EFCA} also exceed 50\% on both metrics. More interestingly, some retrieval results in \cref{tab:sketches-logos-cartoons} are even better than those in \cref{tab:secret-extraction-retrieval}, particularly for sketches and logos. These observations suggest two things. \emph{First}, the attack can remain effective even when the secret images come from different domains. \emph{Second}, secret images with a single, clearly defined semantic concept are easier for the vision model to recognize and retrieve (and, of course, even more so for humans).

\subsection{Attacking Multi-Image Hiding}
\label{subsec:multi-image steganography}
As an important branch of image hiding, multi-image hiding aims to hide multiple secret images within a single cover image~\cite{multi-image-survey}. We explore whether our attack applies to this setting, \emph{i.e.}, whether it can recover recognizable content corresponding to at least one, or even multiple, secret images from a single stego image containing multiple secrets. We select INN-based DeepMIH~\cite{DeepMIH} and autoencoder-based StegFormer~\cite{StegFormer} as representative models. Both are evaluated with two secret images hidden in each cover image. The experiments are conducted on the COCO test dataset.

\begin{figure}[!t]
\begin{center}
\includegraphics[width=7.5cm]{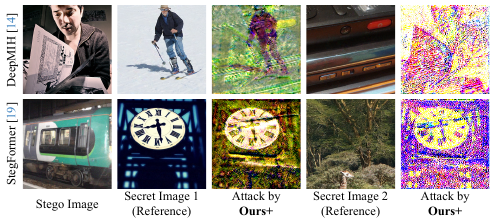}
\caption{\textbf{Visual results of our attack on multi-image hiding.} We use the attack with augmentation (\textbf{+}).}
\label{fig:multi}
\end{center}
\vspace{-1em}
\end{figure}

The visual results are shown in~\cref{fig:multi}. For both models, the core semantics of the first secret image are clearly exposed, while those of the second secret image are not recovered. For StegFormer~\cite{StegFormer}, the second extraction output incorrectly reflects the semantics of the first secret image rather than those of the second. Details of the quantitative results are provided in the supplemental material. Although the attack reveals the semantics of only one secret image, these results show that it remains partially effective against multi-image hiding.

\subsection{Attacking a New Privacy-Preserving Method}
\label{subsec:privacy-application}

\begin{figure}[!t]
\begin{center}
\includegraphics[width=7.5cm]{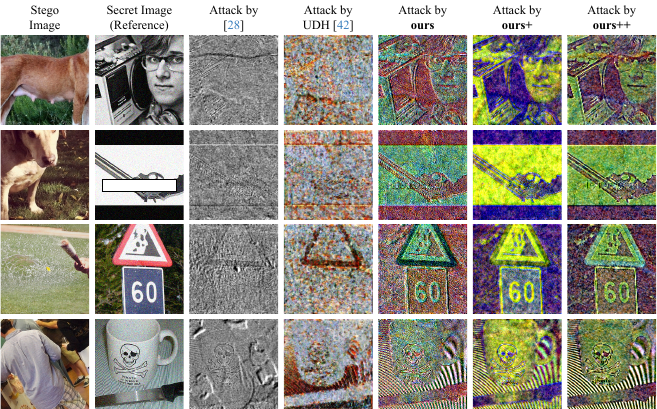}
\caption{\textbf{Visual results of attacks on AIRPNet~\cite{AIRPNet}.} In each row, the last five columns show the corresponding attack outputs.}
\label{fig:attack-airpnet}
\end{center}
\end{figure}

As discussed in~\cref{sec:related work}, an increasing number of studies are exploring image hiding or image-in-image steganography for privacy protection. Can our method attack these applications out of the box? The recently proposed AIRPNet~\cite{AIRPNet}, a privacy-preserving image restoration framework, provides a natural test case. As expected, \cref{fig:attack-airpnet} shows that our attack readily reveals recognizable semantic content from the secret images. This means that anyone equipped with this attack can easily tell what the hidden images depict. \cref{tab:attack-airpnet} further confirms this \emph{privacy vulnerability}. For a more rigorous evaluation, we also compare our attack with UDH~\cite{udh} and the recent steganalysis method of~\cite{possible-steg}, since both studies report that their methods can reveal some secret content. As shown in~\cref{fig:attack-airpnet} and~\cref{tab:attack-airpnet}, the method of~\cite{possible-steg} does reveal rough outlines of the secret images hidden by AIRPNet~\cite{AIRPNet}, although its extraction performance remains limited. The UDH~\cite{udh} outputs contain very little that the vision model can use to identify the original secret image.

\begin{table}[!t]
\caption{\textbf{Secret retrieval for attacks on AIRPNet.} All three variants of our attack outperform~\cite{possible-steg,udh} on COCO and ImageNet.}
  \label{tab:attack-airpnet}
  \definecolor{attackblue}{HTML}{E6F0FA}
  \centering
  \footnotesize
  \setlength{\tabcolsep}{1pt}
  \renewcommand{\arraystretch}{1}
  \setlength{\dashlinedash}{3pt}
  \setlength{\dashlinegap}{3pt}

  \begin{tabularx}{\columnwidth}{
    @{}c*{6}{>{\centering\arraybackslash}X}@{}
  }
    \toprule
    \multirow{4}{*}{Attack}
      & \multicolumn{3}{c}{COCO}
      & \multicolumn{3}{c}{ImageNet} \\
    \cmidrule(lr){2-4}
    \cmidrule(l){5-7}
      & R@1$\uparrow$ & R@10$\uparrow$ & R@20$\uparrow$
      & R@1$\uparrow$ & R@10$\uparrow$ & R@20$\uparrow$ \\
    \midrule
    UDH~\cite{udh}
      & 0.48 & 1.88 & 2.80 & 0.80 & 1.90 & 2.48 \\
    Mallet et al.~\cite{possible-steg}
      & 8.44 & 17.68 & 21.76 & 8.72 & 15.44 & 17.68 \\
    \midrule
    \rowcolor{attackblue}
    \textbf{Ours}
      & 17.00 & 33.40 & 39.76 & 21.00 & 34.88 & 39.50 \\
    \rowcolor{attackblue}
    \textbf{Ours+}
      & 35.60 & 54.24 & 60.88 & 36.56 & 52.62 & 56.74 \\
    \rowcolor{attackblue}
    \textbf{Ours++}
      & 45.00 & 61.60 & 66.72 & 44.92 & 60.20 & 64.54 \\
    \bottomrule
  \end{tabularx}
\end{table}

\subsection{Further Discussions}
\label{subsec:discussions}
While these results are striking, the settings that affect attack performance and the deeper reasons behind this phenomenon deserve further discussion. Following the analysis in~\cref{subsec:privacy-application}, we also include AIRPNet~\cite{AIRPNet} in this section.

\subsubsection*{Effect of augmentation.}

\begin{figure*}[!t]
\begin{center}
\includegraphics[width=16cm]{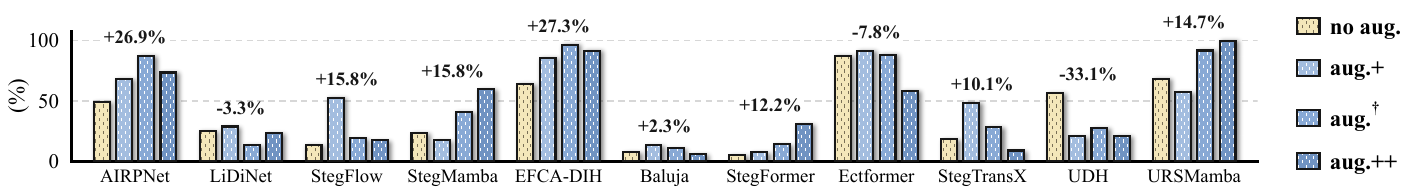}
\caption{\textbf{Effect of training augmentation on secret retrieval.} For each of the 11 target models, we compare \textbf{+}, \textbf{++}, and $\dagger$ with the unaugmented attack and report the average of the three R@50 gains. For $\dagger$, we reduce the augmentation strength by 25\% relative to \textbf{+}.}
\label{fig:augment}
\end{center}
\vspace{-0.5em}
\end{figure*}

In~\cref{tab:secret-extraction-retrieval}, we investigate how our training augmentations affect attack performance and generalization. \cref{fig:augment} further quantifies the gains brought by these augmentations. Specifically, we report the improvements in R@50 over the attack without augmentation across 11 target models, considering the two previously introduced variants (\textbf{+} and \textbf{++}) and a new variant ($\dagger$) with milder augmentation than \textbf{+}. As can be seen, these augmentations yield substantial improvements for the majority of target models. We want to emphasize again that, although these augmentations do not improve attacks on every model (\emph{e.g.}, UDH~\cite{udh}), the gains themselves further support the existence of a shared embedding pattern across most of these models. For the attacker, these gains come at no additional inference cost.

\subsubsection*{Effect of training data.}

\begin{table}[!t]
\centering
\caption{\textbf{Effect of training dataset and size.} We train one extractor on 2,000 ImageNet images and another on 10,000.}
\label{tab:im-coco-2000-10000}
\footnotesize
\setlength{\tabcolsep}{2.5pt}
\renewcommand{\arraystretch}{1}
\begin{tabularx}{\columnwidth}{
    @{}l*{6}{>{\centering\arraybackslash}X}@{}
  }
\toprule
 & \multicolumn{3}{c}{ImageNet-2000} & \multicolumn{3}{c}{ImageNet-10000} \\
\cmidrule(lr){2-4}\cmidrule(lr){5-7}
Method & R@1$\uparrow$ & R@10$\uparrow$ & R@20$\uparrow$ & R@1$\uparrow$ & R@10$\uparrow$ & R@20$\uparrow$ \\
\midrule
HiNet~\cite{HiNet} & 31.08 & 51.12 & 56.16 & 41.00 & 61.48 & 67.12 \\
LiDiNet~\cite{lidinet} & 7.00 & 14.08 & 16.52 & 3.56 & 7.36 & 9.04 \\
StegFlow~\cite{stegflow} & 13.84 & 24.64 & 28.36 & 5.52 & 11.60 & 14.16 \\
StegMamba~\cite{stegmamba} & 10.92 & 22.80 & 26.40 & 15.28 & 27.04 & 31.68 \\
EFCA-DIH~\cite{EFCA} & 33.08 & 53.56 & 59.44 & 23.32 & 40.00 & 46.36 \\
AIRPNet~\cite{AIRPNet} & 8.80 & 18.96 & 23.68 & 9.80 & 20.24 & 25.52 \\
\cdashline{1-7}
Baluja~\cite{baluja2019hiding} & 2.08 & 6.36 & 9.20 & 3.52 & 10.92 & 14.48 \\
StegFormer~\cite{StegFormer} & 2.40 & 6.84 & 8.76 & 1.04 & 3.08 & 4.72 \\
Ectformer~\cite{ectformer} & 64.08 & 79.64 & 82.80 & 77.32 & 89.40 & 91.96 \\
StegTransX~\cite{StegTransX} & 5.12 & 11.52 & 14.40 & 9.60 & 20.04 & 23.44 \\
UDH~\cite{udh} & 12.28 & 24.88 & 29.80 & 15.20 & 32.68 & 39.16 \\
URSMamba~\cite{ursmamba} & 45.72 & 67.48 & 73.28 & 50.44 & 69.92 & 75.28 \\
\bottomrule
\end{tabularx}
\end{table}

\begin{table*}[!t]
\caption{\textbf{Cross-decoder secret retrieval.}
For each decoder (column), we report mean recall (\%) across the 11 other source models (rows).
We start with HiNet~\cite{HiNet}, but most other decoders achieve higher recall. The complete results are provided in the supplementary material.}
  \label{tab:11x11}
  \centering
  \scriptsize
  \setlength{\tabcolsep}{1pt}
  \renewcommand{\arraystretch}{1}

  \resizebox{\textwidth}{!}{%
    \begin{tabular}{@{}l*{11}{c}@{}}
      \toprule
      Metric
        & HiNet~\cite{HiNet}
        & LiDiNet~\cite{lidinet}
        & StegFlow~\cite{stegflow}
        & StegMamba~\cite{stegmamba}
        & EFCA-DIH~\cite{EFCA}
        & Baluja~\cite{baluja2019hiding}
        & StegFormer~\cite{StegFormer}
        & Ectformer~\cite{ectformer}
        & StegTransX~\cite{StegTransX}
        & UDH~\cite{udh}
        & URSMamba~\cite{ursmamba} \\
      \midrule
      R@1$\uparrow$
        & 17.80 & 22.63 & 31.31 & 11.45 & 30.27 & 0.63
        & 28.30 & 7.37 & 35.68 & 1.57 & 25.95 \\
      R@10$\uparrow$
        & 28.96 & 35.39 & 44.59 & 22.43 & 42.40 & 1.70
        & 41.67 & 13.81 & 49.57 & 4.33 & 39.65 \\
      R@20$\uparrow$
        & 32.80 & 39.44 & 48.60 & 26.72 & 46.24 & 2.37
        & 45.63 & 16.39 & 53.97 & 5.70 & 43.96 \\
      \bottomrule
    \end{tabular}%
  }
  \vspace{-0.5em}
\end{table*}

In~\cref{subsec:experimental-setup}, we follow HiNet's official training procedure~\cite{HiNet} and train our secret extractor on DIV2K~\cite{DIV2K}. What if we use a different training dataset? To find out, we randomly sample 2,000 and 10,000 images from ImageNet and train secret extractors on the two subsets separately. As shown in~\cref{tab:im-coco-2000-10000}, replacing the training dataset still yields effective attacks against all the evaluated models. Moreover, increasing the number of training images from 2,000 to 10,000 further improves attack performance against most target models. These results also suggest two things. \emph{First}, an attacker need not worry about which image dataset to use for training the secret extractor. \emph{Second}, an attacker can obtain even stronger extraction results by deliberately using a larger dataset to train the secret extractor.

\subsubsection*{Can a ResNet-based model do this too?}
\label{subsubsec:resnet-attack}
\begin{figure}[!t]
\begin{center}
\includegraphics[width=8cm]{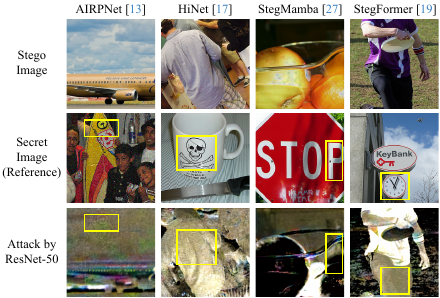}
\caption{\textbf{Visual results of attacks using a ResNet-50 extractor.} The yellow boxes highlight faint traces of secret content.}
\label{fig:attack-by-resnet}
\end{center}
\vspace{-1em}
\end{figure}

Let us consider a more general case: the attacker has never heard of HiNet~\cite{HiNet} and instead uses a standard ResNet~\cite{resnet} to build the hiding and revealing networks. Would the resulting secret extractor still work? Surprisingly, it does too. As shown in~\cref{fig:attack-by-resnet}, when attacking four representative target models, the ResNet-50-based extractor produces outputs that are dominated by contours from the stego images but also contain traces of secret content. Keep in mind that ResNet~\cite{resnet} is just a standard neural network, with no components specifically designed for steganography, yet it can still develop this capability. This makes us wonder: \emph{can other hiding models also achieve the striking extraction results seen with HiNet?} We then conduct cross-model extraction experiments across all evaluated methods. As shown in~\cref{tab:11x11}, extractors from all tested models exhibit this capability to varying degrees, both within and across architectures.

\begin{figure}[!t]
\begin{center}
\includegraphics[width=8cm]{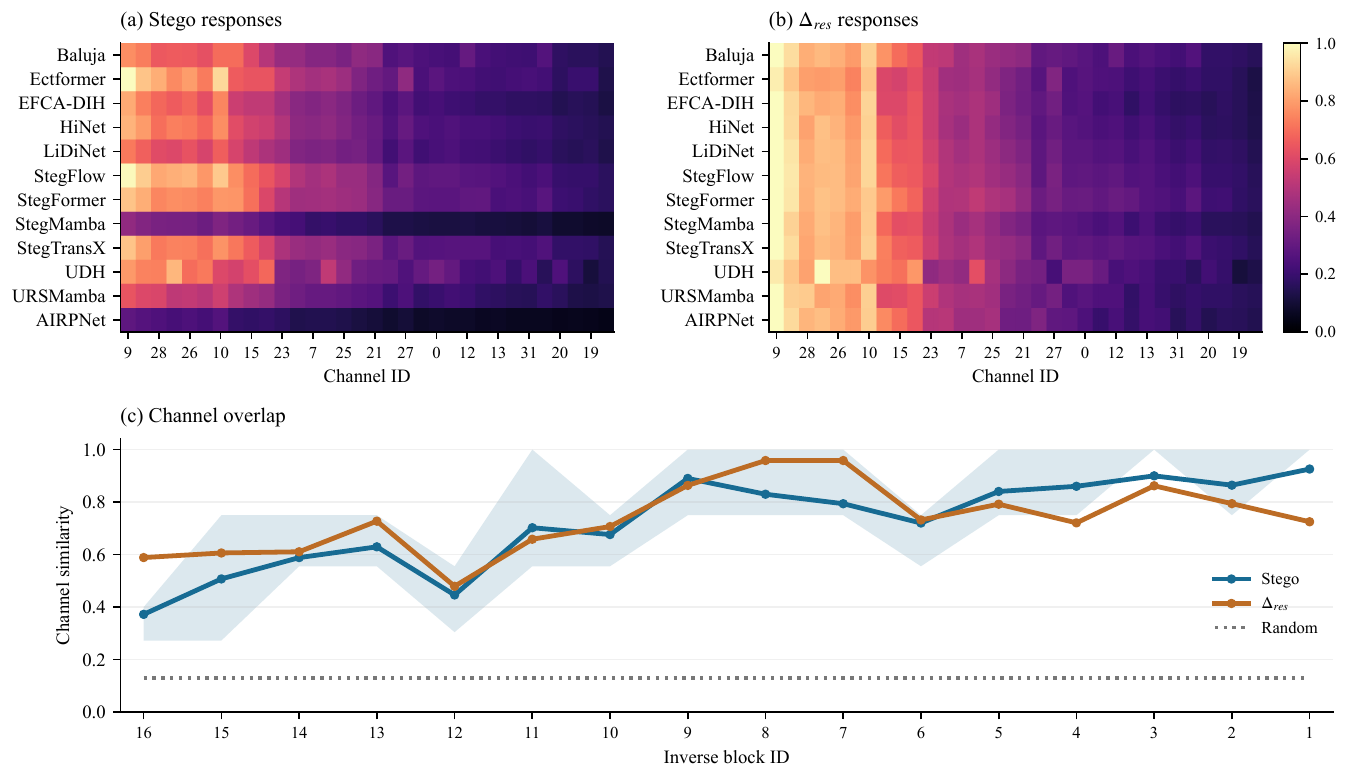}
\caption{\textbf{Shared channel responses in HiNet's decoder.} (a) At block~8, stego images from different models produce the largest activation changes in many of the same channels (brighter cells). (b) This pattern persists when we randomize the Fourier phase of $\Delta_{\mathit{res}}$ and add the residual back to its cover. (c) For both inputs, the mean pairwise Jaccard similarity between sources' top-20\% channel sets exceeds the random baseline in all 16 inverse blocks.}
\label{fig:final_analysis}
\end{center}
\vspace{-0.5em}
\end{figure}

What, then, makes such transfer possible? We probe a fixed HiNet~\cite{HiNet} decoder at the fourth convolution of each inverse block's $\eta$ subnetwork, measuring activation changes relative to matched covers. As shown in~\cref{fig:final_analysis}, in block~8, for example, channels~9 and~28 rank among the most responsive for every stego source. Interestingly, many of the same channels still stand out when we disrupt the residuals' spatial organization while preserving their amplitude spectra. This suggests that the decoder picks up spectral patterns shared by the embedding traces of different hiding models.

These observations suggest an explanation with two connected parts. \emph{First}, end-to-end training may lead the encoder and decoder to process secret information in similar ways. In this sense, the secret extractor learns not only how to decode secret content but also how it is encoded. \emph{Second}, most of the evaluated hiding models exhibit similar embedding preferences. This commonality may make it easier for an extractor trained on one model to recover secret content embedded by others. The situation appears even worse than we initially imagined: each tested model can act as a ``master key'', capable of both hiding its own secret images and revealing secret content hidden by other models.

\section{Conclusion}
\label{sec:conclusion}
In this paper, we show that independently trained decoders can reveal secret semantics from unseen hiding models, with varying effectiveness across architectures. Our analyses suggest that end-to-end training enables decoders to recognize embedding patterns shared across models. However, the precise mechanism remains an open question. The threat also extends to applications that rely on steganography for privacy protection. \emph{From now on, hiding an image using deep image hiding or image-in-image steganography does not guarantee that its content remains private}.

{
    \small
    \bibliographystyle{ieeenat_fullname}
    \bibliography{main}
}



 \end{document}